\documentclass[newstyle,twocolumn,proceedings]{rmaa}

\newcommand{\stas}{Stasi\'{n}ska}

\newcommand{\cmcub}{~cm$^{-3}$}
\newcommand{\kms}{~km~s$^{-1}$}

\newcommand{\Zs}{~Z$_{\odot}$}
\newcommand{\Ms}{~M$_{\odot}$}

\newcommand{\msun}{\ifmmode M_{\odot} \else M$_{\odot}$\fi}
\newcommand{\rsun}{\ifmmode R_{\odot} \else R$_{\odot}$\fi}
\newcommand{\lsun}{\ifmmode L_{\odot} \else L$_{\odot}$\fi}
\newcommand{\zsun}{\ifmmode Z_{\odot} \else Z$_{\odot}$\fi}

\newcommand{\Tstar}{T$^{\star}$}

\newcommand{\Mneb}{M$_{neb}$}

\newcommand{\vexp}{v$_{exp}$}

\newcommand{\Ha}{H$\alpha$}
\newcommand{\Hb}{\ifmmode {\rm H}\beta \else H$\beta$\fi}

\newcommand{\hii}{H~{\sc ii}}

\newcommand{\Hei}{He~{\sc i} $\lambda$5876}

\newcommand{\Ciii}{C~{\sc iii}] $\lambda$1909}

\newcommand{\nii}{[N~{\sc ii}]}

\newcommand{\oii}{[O~{\sc ii}]}
\newcommand{\Oiii}{[O~{\sc iii}] $\lambda$5007}
\newcommand{\Oiiit}{[O~{\sc iii}] $\lambda$4363}
\newcommand{\oiii}{[O~{\sc iii}]}

\newcommand{\rOiii}{[O~{\sc iii}] $\lambda$4363/5007}
\newcommand{\rNii}{[N~{\sc ii}] $\lambda$5755/6584}

\newcommand{\Ostrl}{([O~{\sc iii}] $\lambda$5007 + [O~{\sc ii}] $\lambda$3727) / H$\beta$}

\newcommand{\Np}{N$^{+}$}

\newcommand{\Op}{O$^{+}$}
\newcommand{\Opp}{O$^{++}$}

\newcommand{\Te}{T$_{\rm e}$}

\title{The electron temperature in ionized nebulae}
\author{Gra\.{z}yna Stasi\'{n}ska
  \affil{DAEC, Observatoire de Meudon, 92195 Meudon, France 
(grazyna.stasinska@obspm.fr)}} 

\shortauthor{Stasi\'{n}ska}
\shorttitle{The temperature in ionized nebulae}


\abstract{
We discuss the role of the electron temperature in abundance 
determinations in ionized nebulae (planetary nebulae and 
giant \hii\ regions). We show that, even when observations 
provide a direct estimate of T$_{\rm e}$, abundance determinations may sometimes be 
significantly in error for reasons other than hypothetical temperature 
fluctuations or uncertainties in ionization correction factors, and 
we show striking examples both in the high and the low metallicity 
regimes. On the other hand, even without a direct measure of T$_{\rm e}$ and 
under certain conditions, it is possible to give rough estimates (or 
limits) on the abundances, both in GHRs and in PNe. 
Although the main mechanisms determining the electron temperature in ionized nebulae have 
been known for half a century, we emphasize that there are still important
 problems to be solved. 
 Finally, we advocate for a 
different description of temperature inhomogeneities than the 
 scheme generally used.
}
\resumen{
We discuss the role of the electron temperature in abundance 
determinations in ionized nebulae (planetary nebulae and 
giant \hii\ regions). We show that, even when observations 
provide a direct estimate of T$_{\rm e}$, abundance determinations may sometimes be 
significantly in error for reasons other than hypothetical temperature 
fluctuations or uncertainties in ionization correction factors, and 
we show striking examples both in the high and the low metallicity 
regimes. On the other hand, even without a direct measure of T$_{\rm e}$ and 
under certain conditions, it is possible to give rough estimates (or 
limits) on the abundances, both in GHRs and in PNe. 
Although the main mechanisms determining the electron temperature  in ionized nebulae have 
been known for half a century, we emphasize that there are still important
 problems to be solved. 
 Finally, we advocate for a 
different description of temperature inhomogeneities than the 
 scheme generally used.
}

\listofauthors{G. Stasi\'{n}ska}
\indexauthor{Stasi\'{n}ska, G.}

\begin{document}

\maketitle

\section{Introduction}

Emission lines, and especially the forbidden lines used for abundance 
determinations in ionized nebulae are strongly sensitive to the 
electron temperature, \Te. It is common thought that, when \Te\ can 
be derived directly from observed spectra, for example using the 
\rOiii\ ratio, abundance determinations are fairly reliable. We will 
show that this is not necessarily the case (\S 3). We will also 
discuss the reliability of abundance determinations when \Te\ is not 
measured directly, both in the case of planetary nebulae (PNe) and giant \hii\ 
regions (GHRs) (\S 4). Then, we will 
address the question of whether temperatures observed in ionized nebulae
 are compatible with the predictions from photoionization 
models and we will touch upon the problem of temperature inhomogeneities (\S 5).

\section{Basics}

As is known from general considerations on the thermal balance in 
ionized nebulae 
(Spitzer 1948, 1949, Spitzer \& Savedoff 1950, Osterbrock 1989), 
the electron temperature results from a balance between energy gains and losses.
 It is easy to show that, schematically, the energy gains per unit 
volume per unit time in a photoionized nebula can be written as:
\[G =n_{e}^2 \alpha _{A}(H^{0}, T_{e}) \frac{3}{2} \overline E  \]

where $n_{e}$ is the electron density, $\alpha _{A}(H^{0}, T)$ is the 
recombination coefficient of hydrogen, and $\overline E$ is the mean 
energy of the absorbed photoelectrons,
 which is roughly rooportional to the effective 
temperature of the ionizing radiation field, \Tstar.
 Energy losses are due to 
a variety of processes, emission of collisionally excited lines 
 being the most important. In a two-level approximation and neglecting collisional deexcitation,  
they can be schematically written 
as:
\[L = \sum_{ijl} n_{e} n(X^{ij}) 
\frac{\Omega_l}{\omega_l}
T_{e}^{-0.5} e^{-\chi_{l}/kT_{e}}  h\nu_{l} \]
where $n(X^{ij})$ is the number density of ion $X^{ij}$, and  the index $l$ refers to a specific line, 
$\Omega_{l}$ being the collision strenght, 
and $\chi_{l}$ the excitation level. 

Thus, the electron temperature in a nebula is not a strong function of
 the distance to the ionizing stars, since it  does not depend on the flux of ionizing photons.
It is larger for higher values of \Tstar\ and lower values of the metallicity, $Z$.
Fig. 1 illustrates the heating and cooling rates in a photoionized nebula, at two different densities.
At high density, the infrared fine-structure lines are partly quenched by collisional 
deexcitation, resulting in a larger \Te.
Note also that, because of the shape of the cooling curves, 
there is a large temperature difference between the high and the low excitation
 zones at high metallicities. 

\begin{figure}[h]
\begin{center}
\includegraphics[width=\columnwidth]{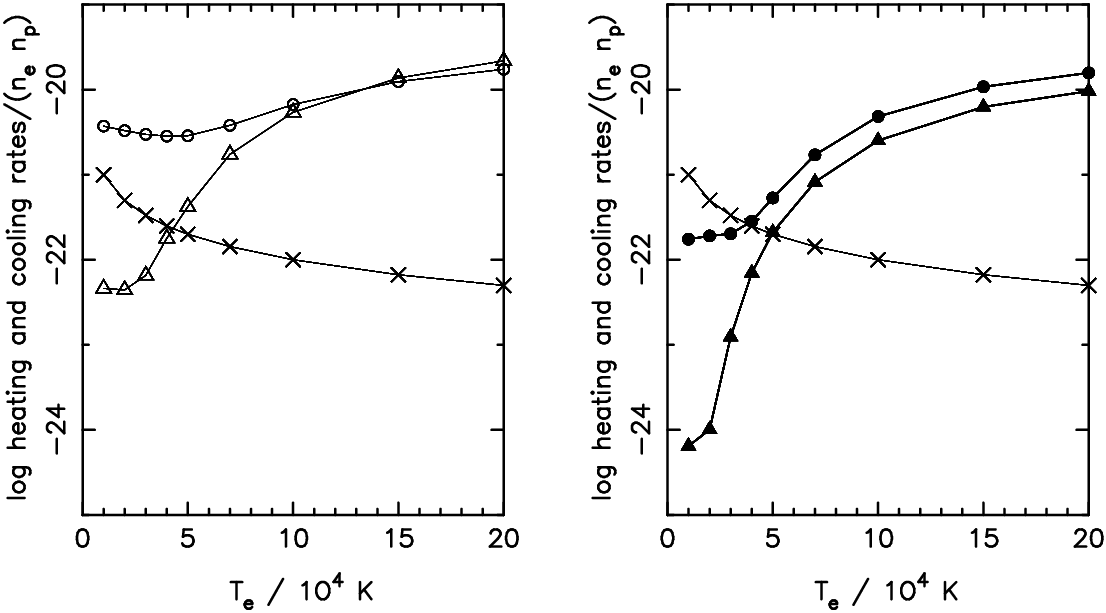}
  \caption{Approximate heating and cooling rates divided by 
$n_{e}^{2}$, in arbitrary units. Circles: approximate cooling rates in the high 
excitation zone (\Opp\ ions only). Triangles: cooling rates in the 
low excitation zones (\Op\ and \Np\ ions only). Left: 
$n_{e}$=100\cmcub;  right: $n_{e}$=10\,000\cmcub.  The heating rate is 
represented by crosses.
 By shifting the heating curve with 
respect to the cooling curves, one can understand how \Te\ varies with 
metallicity or with the effective temperature of the ionizing 
radiation field. 
}
\end{center}
  \label{fig:1}
\end{figure}

\section{When \Te-based abundance determinations are in error}

\subsection{One example at low metallicity: the effect of collisional 
excitation of \Ha }

\begin{figure}[h]
\begin{center}
\includegraphics[width=\columnwidth]{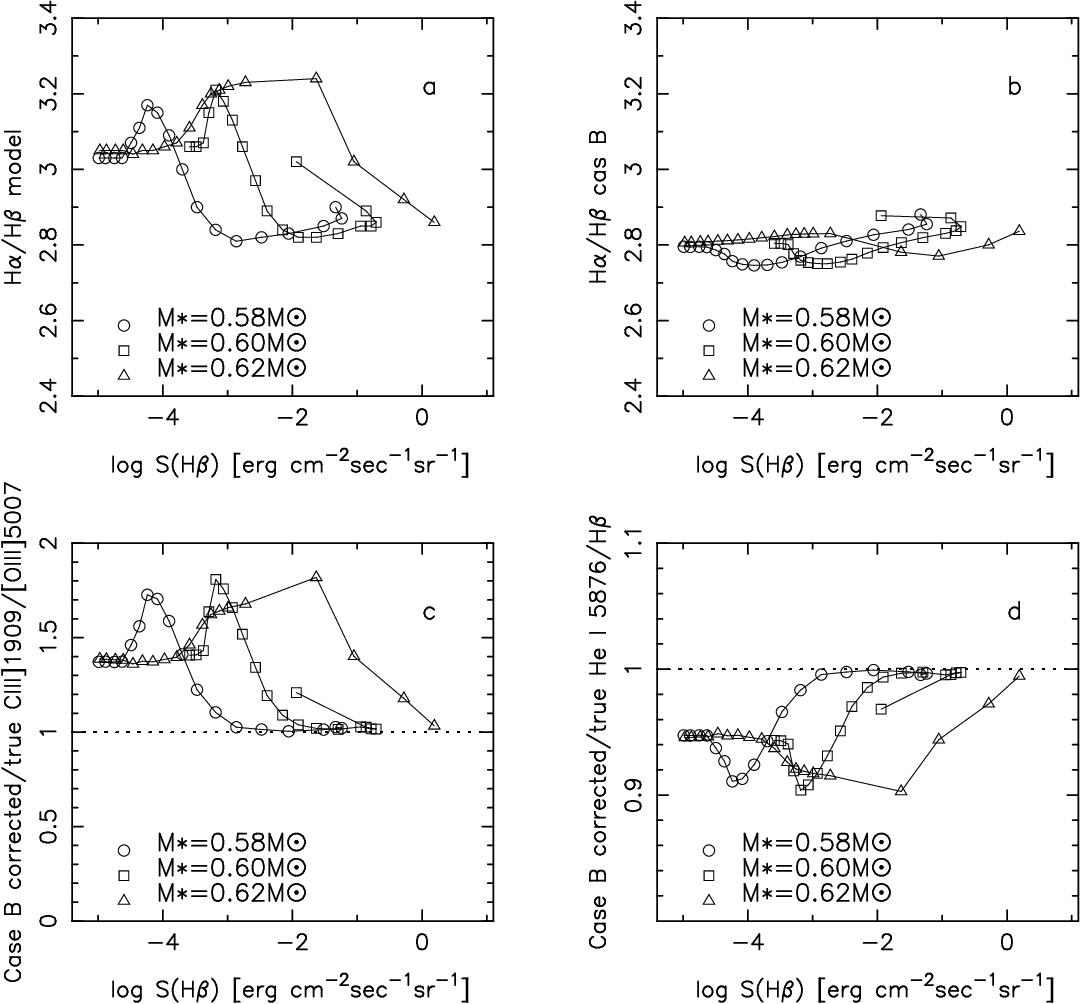}
  \caption{Photoionization models for evolving PNe with $Z$ = 
0.1~\Zs, 
 \Mneb=0.2\Ms, \vexp=20\kms. 
a) computed \Ha/\Hb;
 b) case B  \Ha/\Hb; 
c) \Ciii/\Oiii\ ``dereddened'' using case B with respect to actual 
\Ciii/\Oiii; 
d) same as c) for \Hei/\Hb. }
\end{center}
  \label{fig:2}
\end{figure}

At low metallicities, because of the high \Te, collisional  excitation of the \Ha\ 
line becomes non negligible and the \Ha/\Hb\ ratio no longer has the recombination value. 
If the dereddening procedure of an observed spectrum involves \Ha\ 
(which is commonly the case), this translates into an overestimation 
of the extinction C(\Hb), and an error in the dereddened line ratios. 
It can easily be shown that this error in the line ratios  is  \emph{independent of extinction}.
It can be large 
when the lines in question have very different wavelenghts.

This is illustrated in Fig. 2, which shows the results of 
photoionization models at metallicity $Z$=  0.1~\Zs\ for evolutionary 
sequences of planetary nebulae whose central stars follow
 the theoretical tracks of   Bl{\"{o}}cker (1995).  These 
models were computed using the procedure outlined in Stasi\'{n}ska et al (1998)
\footnote{Unless otherwise stated, the photoionization calculations presented here used the code 
PHOTO as described by  Stasi\'{n}ska \& Leitherer (1996).}. 
We see that, at such a metallicity, there is quite a range in epochs 
during the evolution of PNe (materialized in the figure by the surface brightness in \Hb), 
where \Ha/\Hb\ is significantly larger than the recombination value. 
The 
resulting error in the derived  helium abundance may reach 5 -- 10\%.
 A similar effect is expected in very low metallicity \hii\ regions. 
This is important to keep in mind for 
the determination of the pregalactic helium abundance from low metallicity GHRs. 
Davidson \& Kinman 
(1985) indeed warned about this effect, but it does not seem to have been properly 
taken into account in later studies (see e.g. Olive et al. 1997 or 
Izotov et al. 1999 and references therein). The error in  C/O 
obtained using the \Ciii\ and \Oiii\ lines may reach almost a factor 2 in the 
PN models shown here. Again, a similar effect is expected in low 
metallicity GHRs. This could explain why the carbon abundance 
estimated in  I Zw 18 is larger than expected for its metallicity 
(Garnett et al 1997a). However, in the case of UV data, a possibly even more important problem 
is the uncertainty in the reddening law (Garnett et al 1999).

\subsection{One example at high metallicity: the absence of \oiii\
optical lines from the \Opp\ zone}

At high $Z$, because of the intense cooling by the forbidden lines, 
\Te\ is low (see Fig. 1) and the transauroral lines used for \Te\ diagnostics 
become hard to detect. So far, there has been no \Te-based 
measurement of metallicities higher than about 0.5~\Zs\ in any GHR. 
With 8~m-class telescopes, however, it becomes possible to detect 
weak \Oiiit\ fluxes, and the hope has been expressed that 
it will be possible to directly probe oxygen abundances at 
metallicities higher than solar. However, as already shown by \stas\ 
(1978) and Garnett (1992), high metallicities induce an important temperature 
drop in the \Opp\ zone  because cooling is dominated by the 
[OIII]52$\mu$ and [OIII]88$\mu$ lines which are almost independent of \Te\ (see Fig. 1). Therefore, the value of \Te\ derived from  
\rOiii\  would largely overestimate the temperature characteristic of 
the \Opp\ zone. Consequently, classical \Te-based methods would 
strongly underestimate the oxygen abundance. This is illustrated in 
Fig. 3. The derived O/H ratio always stays below the solar value, and 
would be underestimated by orders of magnitudes for metallicities 
over twice solar if such methods were applied. Incidentally, note the
 rise in the \Oiiit\ flux at log O/H + 12 $>$ 9.1. 
This is due to the fact that recombination 
becomes the dominant process for the excitation of this line at the low temperatures involved.

 \begin{figure} [h]
\begin{center}
\includegraphics[width=\columnwidth]{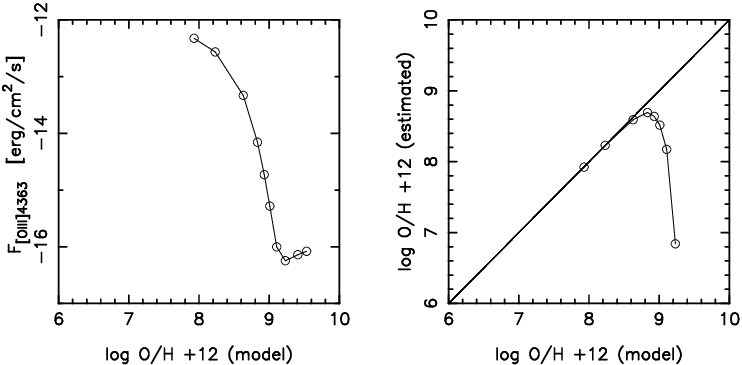}
  \caption{Results from a simple model for a GHR having 
$T_{*}$=45\,000~K, $Q_{H}$=3~10$^{51}$~ph 
s$^{-1}$, $n$=30 \cmcub, filling factor = 0.1.
Left: predicted flux in  the \Oiiit\ line for an object at a distance of 1 Mpc; right: O/H derived by the 
classical $T_{e}$-based method
as a function of input O/H. }
\end{center}
  \label{fig:3}
\end{figure}

\begin{figure} [h]
\begin{center}
\includegraphics[width=\columnwidth]{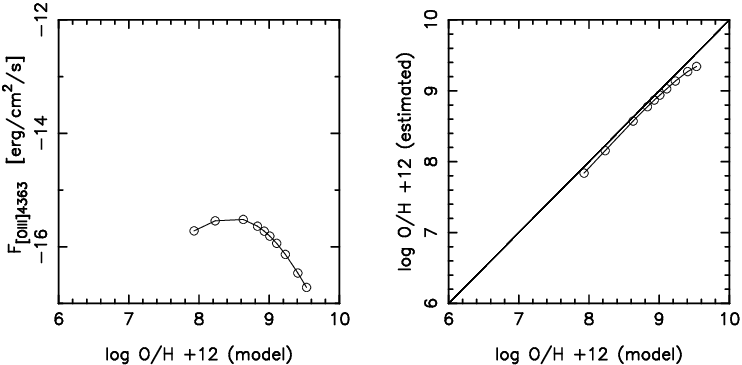}
  \caption{Same as Fig. 3 for a simple model ofa ionization bounded PN  having $T_{*}$=150\,000~K, 
$Q_{H}$=3~10$^{47}$~ph~s$^{-1}$, $n$=10$^{4}$ \cmcub, filling factor 
= 0.3 }
\end{center}
  \label{fig:4}
\end{figure}

Does this mean that it is impossible to reliably determine the oxygen 
abundance in the central parts of galaxies, where it is supposed to be oversolar?
 The most straightforward 
way, of course, would be to use the [OIII]52$\mu$ line (Simpson et 
al. 1995, Rudolph et al. 1997). Another possibility, suggested by \stas\ (1980), would be 
through high spatial resolution emission line  imaging.  Metal rich 
GHRs would show very characteristic and unusual  spatial  
correlations between line intensities, since \oiii\ and \oii\ and 
would be emitted almost cospatially in the high \Te\ zones, while the 
hydrogen and helium recombination lines would preferentially arise in 
the low \Te\ zones. Of course, detailed photoionization modelling 
reproducing the observed tendencies would still be necessary to 
provide a quantitative estimate of O/H. 

Planetary nebulae provide an interesting substitute to GHRs for abundance determinations in the 
central parts of galaxies. 
As shown by Richer et al. (1998), the most 
 luminous PNe are good tracers of the ISM oxygen 
 abundance. Such PNe, being young (therefore dense) and excited by 
stars with \Tstar\ of the order of  100\,000~K, 
have larger \Te\  than GHRs of same metallicity (see Fig. 1).
Therefore, they can be detected in \Oiiit, even though their intrinsic luminosity 
is orders of magnitudes smaller than that of GHRs (compare Figs. 4 and 3). 
In addition, electron temperature gradients are not expected to be so severe as in GHRs
(as can be seen from Fig. 1), 
so that the classical \Te-based method to derive O/H should give reasonable 
 results. Indeed, Fig. 4 shows that, at least theoretically, this is 
the case.

\section{Estimating O/H without \Te}

\subsection{Strong line methods for GHRs and PNe}

It has been shown by Pagel et al. (1979) that, in GHRs, provided that the 
hardness of the ionizing radiation field is related to the 
metallicity, it should be possible to calibrate some well chosen  strong 
line ratios as a function of O/H (see \stas\ 2001 for a summary of the basics and a short 
bibliography of strong line methods). The extremely tight correlation between 
 \Ostrl\ and \Te-based O/H determinations in high 
signal-to-noise observations of \hii\ galaxies (see Fig. 5) 
is an empirical confirmation that this method works at low metallicities, 
as was suggested by Skillman (1998). Incidentally, this implies that 
GHRs in blue compact galaxies constitute a very homogeneous class as 
regards their excitation and nebular properties (see discussion in 
\stas\ et al 2001).
\begin{figure}[h]
\begin{center}
\includegraphics[width=3.7cm]{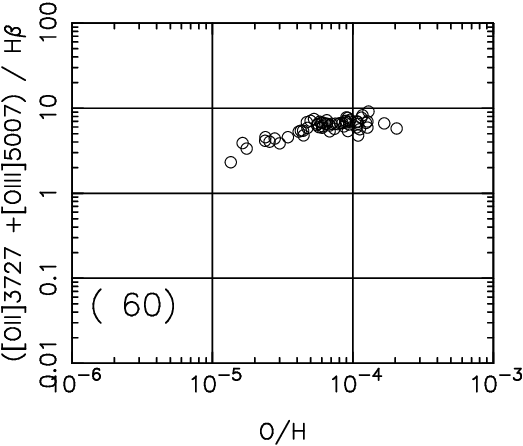}
  \caption{ \Ostrl\ versus O/H for GHRs in a sample of blue 
compact galaxies with \Te-based derivations of O/H. Data from Izotov 
\& Thuan (1999 and references therein). The number of objects is indicated in parenthesis.}
\end{center}
  \label{fig:5}
\end{figure}

\begin{figure}[h]
\begin{center}
\includegraphics[width=\columnwidth]{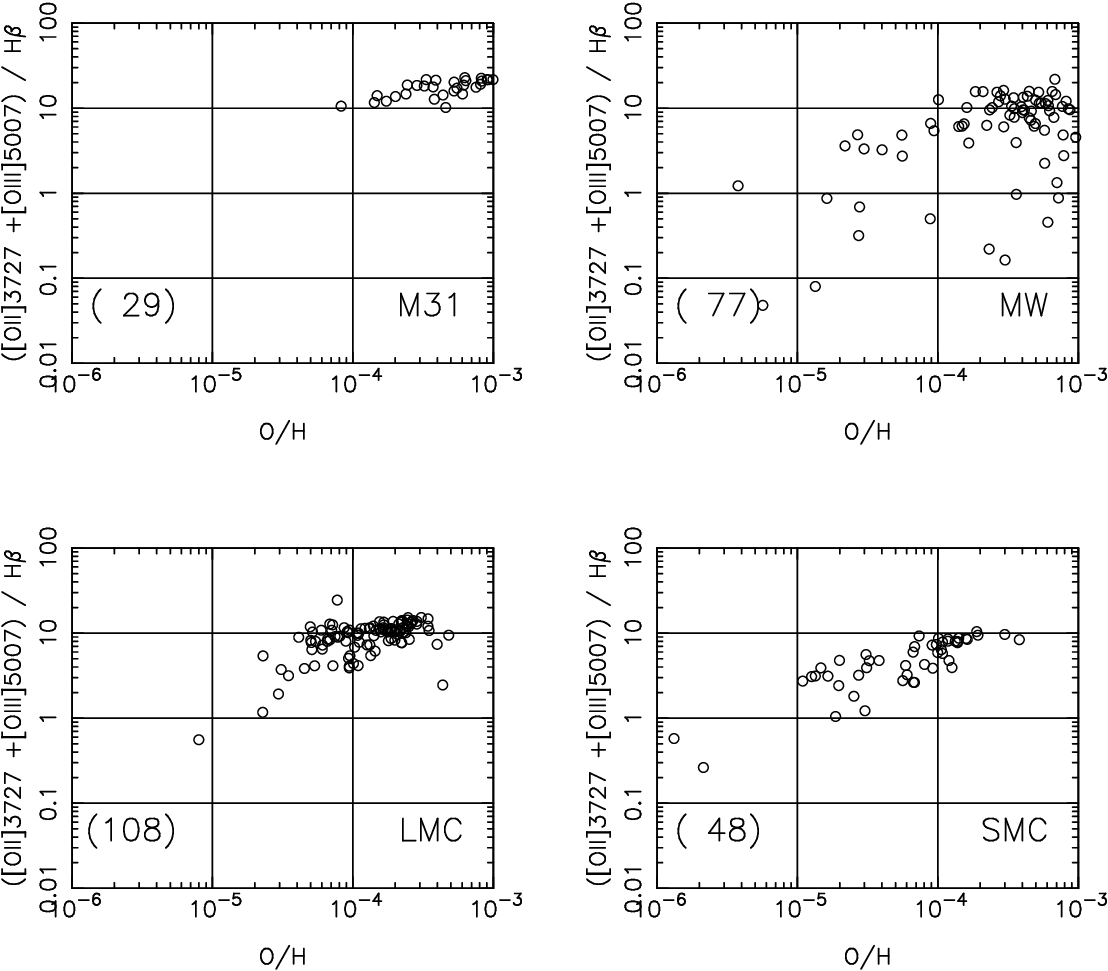}
  \caption{Same as Fig. 5 for PNe in various galaxies: M31, the bulge 
of the Milky Way, the LMC and the SMC. Data as collected in \stas\ et 
al. (1998).}
\end{center}
  \label{fig:6}
\end{figure}

Can such a strong line method be applied to PNe?  Since PN central 
stars span a wide range in \Tstar\ (30\,000~ K -- 200\,000~K), one 
does not expect such methods to work in the same way as for GHRs. 
However, Fig. 6 shows that PNe outline a rather well defined upper envelope
 in the \Ostrl\ vs. O/H plane, as already noted by Richer 
(1993) for Magellanic Clouds PNe. This upper envelope probably 
corresponds to the hottest central stars, and can be used to infer a 
lower limit for O/H. Such a procedure has been used by Richer \& Mc 
Call (1995) and Richer et al. (1998) for extragalactic PNe with 
\Oiiit\ undetected.

\subsection{Photoionization modelling}

Contrary to a widespread opinion, photoionization models (be it of PNe or 
of GHRs) are usually incapable of providing abundances if \Te\ is not 
directly measured.
It is easy to produce models differing only in abundance ratios that 
will reproduce all the observed emission lines. This is illustrated 
in Fig. 7, which shows the run of \oiii/\Hb\ and \oii/\Hb\ as a 
function of O/H for a simple PN model. The reason of such a 
behavior is the same for PNe and GHRs and has been wideley commented 
in the literature. As O/H increases, \Te\ decreases due to enhanced
cooling, and the optical forbidden lines are gradually overtaken by the 
infrared lines, leading to these bell-shaped curves. The maximum of the curves occurs at a metallicity 
around solar for dense PNe ionized by stars with \Tstar\ $>$ 
100\,000~K, and around half solar for GHRs. 

\begin{figure}[h]
\begin{center}
\includegraphics[width=\columnwidth]{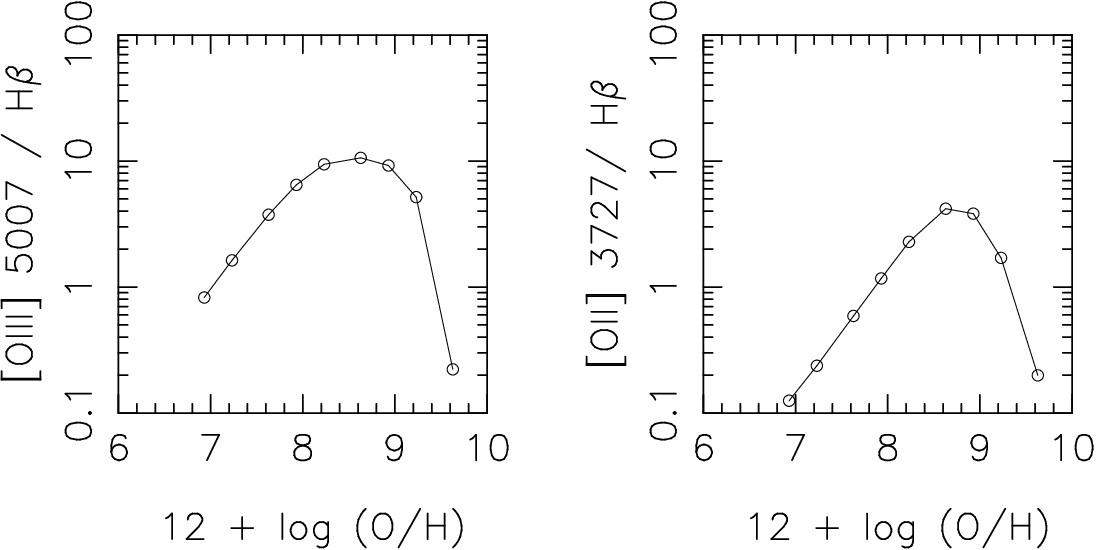}
  \caption{Variation of \oiii/\Hb\ and \oii/\Hb\ as a
 function of O/H for a simple density bounded  model of PN with 
  \Mneb=0.5~\Ms, 
$T_{*}$=100\,000~K, 
$Q_{H}$=3~10$^{47}$ ph s$^{-1}$, $n$=10$^{3}$ \cmcub.}
\end{center}
  \label{fig:7}
\end{figure}

Table 1 shows the results of photoionization modelling of the 
Galactic Bulge PN  PNG 351.2+05.2 (M 2-5) for which Ratag (1991, 
1997) had claimed an abundance of one fourth solar. Using the Lviv photoionization code, \stas, Malkov \& 
Golovaty (1995, unpublished)
 produced two  classes of models for this 
object accounting satisfactorily for the observed data (line 
strenghts, total flux and radius). The first one indeed has an oxygen 
abundance similar to Ratag's, but the second has O/H larger than 
solar!

\begin{table}[h]
  \caption{Models for the PN M 2-5}
  \begin{center}
    \begin{tabular}{p{1.8cm}p{1.1cm}p{1.1cm}p{1.1cm}p{1.1cm}}
\hline\hline
 & obs     &     Ratag   &  SMG low Z      &    SMG high Z       \\ 
\hline
 T* (K)    & & 37500  &  39000 &  39000 \\
 r* (10$^{10}$cm) & &                          &   5.0 &     4.9 \\
 R$_{in}$  (pc)    & &                          &   0.062  &   0.065  
\\
 R$_{out}$ (pc)   &        0.10    &          & 0.087  &  0.085  \\
-log\,F(\Hb) &      11.409      &         & 11.407  &   11.411 \\
\hline
 n (cm$^{-3}$) &            &       2050  &    1800    &   1800 \\
 He        &            &      0.117  &   0.117    &  0.100 \\
 C         &            &             &  1.50(-3    & 1.20(-3 \\
 N         &            &     4.80(-4  &  2.50(-4    & 6.00(-4 \\
 O         &            &     2.20(-4  &  2.40(-4    & 1.20(-3 \\
 Ne        &            &             &  5.00(-5    & 2.40(-4 \\
 S         &            &     2.30(-5  &  3.00(-6    & 7.00(-6 \\
 \hline
 OII 3727  &  0.596  &   &  0.587    &  0.604 \\
 Ne 3869  &   &   &    0.014  &   0.0096 \\
 OIII 4363   & $<$.0013 &          &    0.0006 &   0.0001 \\
   HeII 4686  &         &           &   0.0004 &  0.0003 \\
     HI 4861  &  1.00   &           &   1.00   &  1.00 \\
 OIII 5007  &  0.283  &           &   0.304   &  0.275 \\ 
   NI 5200  &  0.0149 &           &   0.0043  &  0.0087 \\
  NII 5755  &  0.0071: &          &    0.0151  &  0.0060 \\
    HeI 5876  &  0.128  &            &  0.126    &   0.128 \\
   OI 6300  &          &          &   0.0054  &   0.0116 \\
  NII 6584  &  2.85    &          &   2.79    &   2.81 \\
  SII 6717  &  0.0565   &         &   0.053   &  0.0558 \\
  SII 6731  &  0.084    &         &   0.077   &  0.0826 \\
  OII 7325  &  0.0091:  &         &   0.0126  &  0.0063 \\ 

\hline\hline

    \end{tabular}
    \label{tab:1}
  \end{center}
\end{table}

This demonstrates that, in absence of a direct measure of \Te,  
photoionization modelling by itself does not necessarily give a clear 
clue to the metallicity of nebulae when only optical  
data are available. In some cases, however, when  solutions on either 
the low or high metallicity side can be rejected using  
arguments of astrophysical nature, photoionization modelling can provide estimates of O/H. 

An example where photoionization modelling gives interesting limites on O/H 
 is the recently discovered  PN in the galactic halo, SBS 1150+599A 
(now PN G135.9+55.9, for full details see Tovmassian et al., 2001 
and these proceedings). In that case, 
solutions at high metallicity can obviously be discarded. There is 
still a wide range of solutions compatible with the observed 
line ratios, but when additional considerations are taken into 
account, it is found that the object has 
 O/H of the order 1/500 
solar (making it possibly the most oxygen-poor object known in the 
Galaxy).

\section{Do we understand the temperature in ionized nebulae? }

So far, we have not questionned the electron temperature itself. As a matter of fact, 
there are indications that simple photoionization models are incapable of reproducing 
satisfactorily the observed \Te\ in quite a variety of situations. There 
are cases of GHRs as well as of PNe, where photoionization models 
give too low a \Te\ (Campbell 1990, Garc\'{\i}a-Vargas et al. 1997, \stas\ 
\& Schaerer 1999, Luridiana et al. 1999, Luridiana \& Peimbert 2001, 
Pe\~{n}a et al 1998), calling for additional heating sources. In other cases, however, no \Te\ problem is 
apparent (Gonz\'{a}lez Delgado \& P\'{e}rez 2000) and in some cases, 
models seem to give too high a \Te\ (Oey et al. 2000). 
So far, we are still lacking a synthetic view of these anomalies. 
 One difficulty is that, even if a model 
reproduces the observed line ratios satisfactorily, this does not necessarily mean 
that the model is a fair representation of the object. For example, 
the first attempts to solve the \Te\ problem in I Zw 18 were not 
supported by later observations (see discussion in \stas\ \& Schaerer 
1999).

It is of prime importance to understand what determines \Te\ in real 
nebulae if one wishes to derive reliable abundances, 
since one way or another, the observed line ratios used to derive 
abundances from optical or ultraviolet data depend on \Te. 

In the following, we comment on two disputed topics.

\subsection{Temperature gradients in photoionized nebulae}

While the electron temperature, being determined by the hardness of the ionizing photons
and not on their flux (c.f. \S 1), 
 does not necessarily fall off with radius, some 
\Te\ gradients are nevertheless expected. A consideration of the heating and 
cooling curves (Fig. 1) shows that \Te\
should rise outwards if the outer zones contain ions that are less 
efficient for the cooling. As mentioned above, the effect can be quite
strong at metallicity above solar. At low metallicity, on the other 
hand, \Te\ decreases outwards. The hardening of the absorbed photons 
towards the ionization front, due to the frequency dependence of the 
photoabsorption cross-section is another factor that affects the 
radial \Te\ distribution. The predicted effect is however rather mild 
(see e.g. the GHRs models of \stas\ \& Leitherer 1996, or the grid of 
PN models from \stas\ et al. 1998). 

Temperature gradients can be estimated from the comparison of \Te\ 
derived from various line ratios, such as \rOiii\ and \rNii\ (see eg. 
Garnett et al. 1997b for GHRs and Mc Kenna et al. 1996, Kinsburgh \& 
Barlow 1994 for PNe). Recently, Pe\~{n}a et al. (2001) found that in 
PNe with Wolf-Rayet central stars, \Te(\oiii)/\Te(\nii) could be 
significantly different from 1 (by as much as 40\%) and that it 
seemed to decrease with increasing \Opp/\Op. Such a tendency cannot 
be reproduced by classical photoionization models. 
Dopita \& Sutherland (2000) have shown that, by including the 
photoelectric effect on dust grains, one obtains larger \Te\ 
gradients than in dust free models, because dust heating affects essentially the inner zones. 
The cases with \Te(\oiii)/\Te(\nii) $\ll$ 1 observed by Pe\~{n}a et 
al. (2001) however remain unexplained. One possibility is to advocate 
a contribution from shock heating in the outer zones,
 but quantitative models have still to be worked out.

\subsection{Temperature inhomogeneities}

In a seminal paper, temperature inhomogeneities have been advocated by Peimbert (1967) to 
explain the discrepancies between \Te\ as determined by various 
methods. Peimbert introduced the temperature fluctuation scheme and 
observational work accumulated over the years by Silvia Torres-Peimbert, 
Manuel Peimbert and their coworkers and
disciples seems to indicate a 
value of $t^{2} \sim 0.04$. There are many questions about the 
reality, the ubiquity and the causes of these \Te\ fluctuations (see 
e.g. papers by Esteban, Peimbert, Binette, Liu and P\'{e}quignot in 
this volume, and references therein). 

In fact, the temperature fluctuation scheme may not be quite 
appropriate to describe temperature inhomogeneities in ionized 
nebulae. A different approach was presented by Mathis et al. (1998). 
Here, and in order to help visualising the problem, we use a toy 
model consisting of two homogeneous zones of volumes $V_{1}$ and 
$V_{2}$ with  temperatures $T_{1}$ and $T_{2}$, electron densities 
$n_{1}$ and $n_{2}$, and densities of the emitting ions (e.g. \Opp) 
$N_{1}$ and $N_{2}$. If we call $f$ the ratio  $(N_{2} n_{2} V_{2})/(N_{1} n_{1} V_{1})$ 
of the weights of the 
emitting regions, the mean 
electron temperature defined by Peimbert (1967) becomes
\[T_{0}= \frac{ T_{1} + f T_{2}} {1+f} \] and the fluctuation 
parameter becomes
 \[t^{2}= \frac{ (T_{1}-T_{0})^2  + f (T_{2}-T_{0})^2} {(1+f)T_{0}^2} 
\]

\begin{figure}[h]
\begin{center}
\includegraphics[width=\columnwidth]{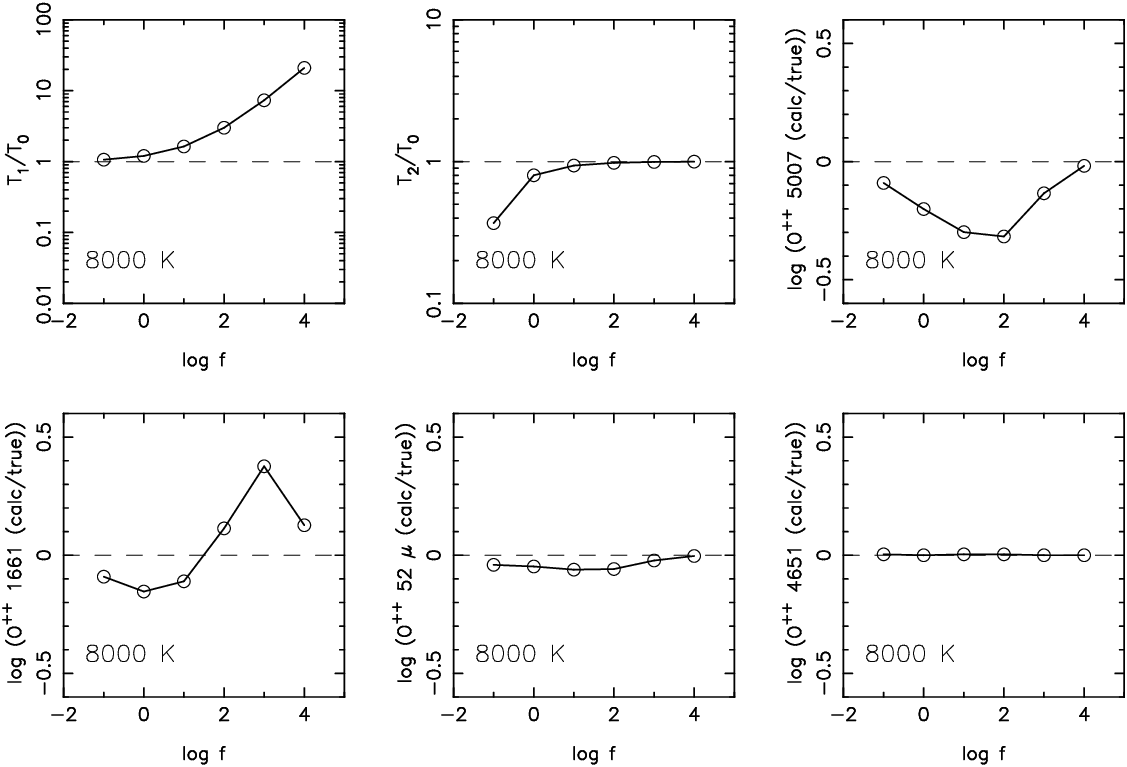}
  \caption{Results from a two zone toy model defined by  $ 
T_{0}$=8\,000~K and $t^{2} = 0.04$ (see text).}
\end{center}
  \label{fig:8}
\end{figure}

\begin{figure}[h]
\begin{center}
\includegraphics[width=\columnwidth]{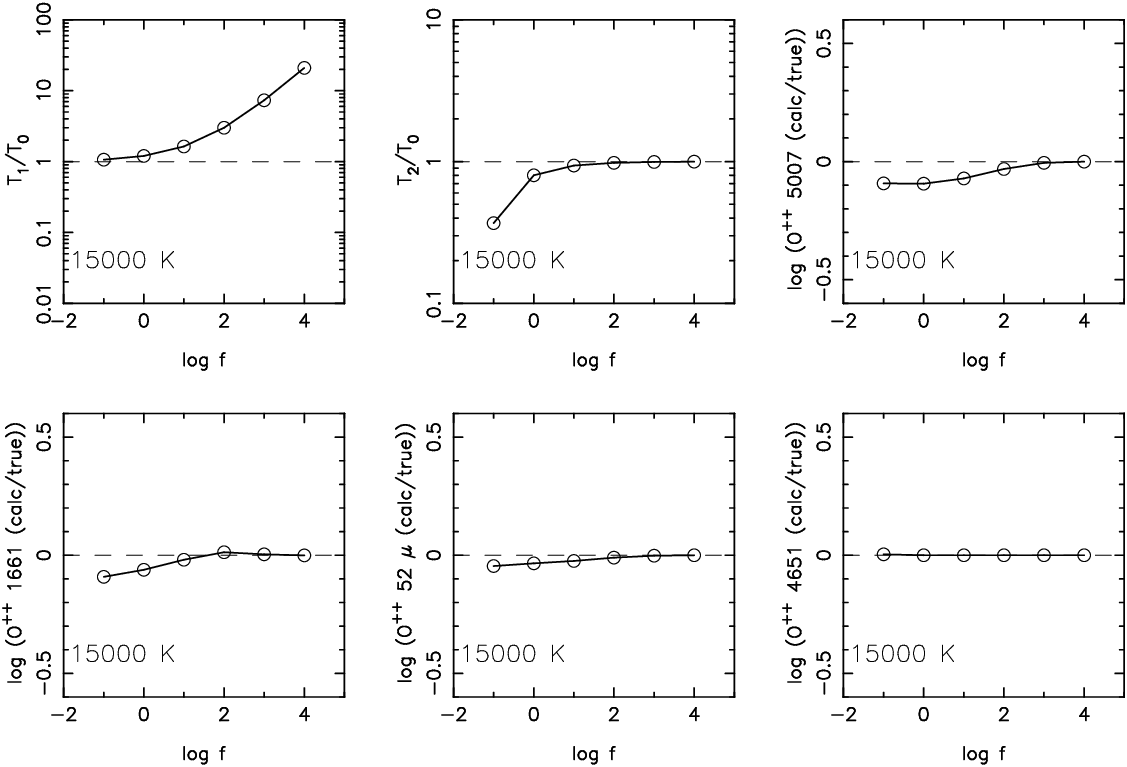}
  \caption{Results from a two zone toy model defined by  $ 
T_{0}$=15\,000~K and $t^{2} = 0.04$ (see text).}
\end{center}
  \label{fig:9}
\end{figure}

In Figs. 8 (resp.  9) we explore the values of $f$ that lead to the 
canonical value $t^{2} = 0.04$ at $T_{0}$=8\,000~K (resp. $ 
T_{0}$=15\,000~K), and show the effect on the  
\Opp\ abundances calculated by using the \Te\ derived from \rOiii. 
The case $f=1$, i.e. regions of equal weight, corresponds to 
$T_{1}$=12\,000~K and $T_{2}=$8\,000~K. This is quite a large temperature 
difference, which would correspond to  heating- or cooling rates 
differing by about a factor 3  between the two zones (see Fig. 1). No 
wonder that photoionization models have difficulties in producing 
such values of $t^{2}$ around these temperatures!
When $f \gg 1$, there is a high weight zone at $T_{2} \le T_{0}$ and 
a low weight zone at $T_{1} \gg T_{0}$. Such a situation could correspond to 
a photoionized nebula with small regions being heted by shocks or conduction. 
When $f \ll 1$, there is a high 
weight zone at $T_{1} \ge T_{0}$ and a low weight zone at $T_{2} \ll 
T_{0}$ which could correspond to high metallicity clumps. 

As expected, Figs. 8 and 9 show that the  \Opp\ derived from infrared fine structure lines
 is correct. The optical recombination line {O~{\sc ii} $\lambda$4651} also give a correct \Opp. 
However, high temperature dielectronic recombinations are not included in 
the present day atomic calculations. They might well boost  
{O~{\sc ii} $\lambda$4651} in zones of high \Te\ (around 20\,000--50\,000~K, see 
Liu et al. 2000, Dinerstein et al. 2000). This is an attractive 
explanation to the fact that in PNe, \Opp\ abundances derived from recombination lines 
are far greater than derived from optical or infrared lines (see Liu, these proceedings).
Compared to the alternate explanation invoking oxygen-rich condensations 
(Liu et al 2000, P\'equignot, these proceedings), this one has the advantage 
of being naturally fulfilled if additional heating mechanisms substantially 
raise the electron temperature in limited volumes of the emitting gas, 
such as shock or conduction fronts. Besides, they do not require 
to find a mechanism to produce oxygen-rich pockets in planetary nebulae.

Following expectations, \Opp\ 
derived from \Oiii\ is generally underestimated, but it is 
interesting to note that the magnitude of the effect depends both $f$ 
and on $T_{0}$. It is negligible when log $f >$ 3--4, because \Oiiit\ 
saturates above $\sim$ 50\,000K. In the $T_{0}$ = 
15\,000~K case, the effect is never strong anyway. 
For the $T_{0}$ = 8\,000K case, errors in \Opp\ of up to a factor 2--3 can 
occur in the regime 
where \Oiiit  is significantly emitted in both zones. 
As for 
{O~{\sc iii}] $\lambda$1661}, it  may lead either to an underestimation
 or to an overestimation of 
\Opp/H, because the dynamical range of the line is larger than that 
of \Oiiit.

Fig. 8 and 9 thus amply demonstrate that the classical temperature 
fluctuation scheme can be misleading. Even in a simple two zone 
model, the situation needs 3 parameteres to be described, not 2. 
In our representation, these parameters would be $T_{1}$, $T_{2}$ and $f$, 
but other definitions can be used. This means that care must be taken 
when drawing conclusions on temperature inhomogeneities by comparing 
abundances derived from various lines. We argue that such diagrams as 
shown here might be useful to pinpoint the physical cause of 
temperature inhomogeneities in ionized nebulae.

\acknowledgments

This wandering - and wondering - about the electron temperature in 
ionized nebulae is dedicated to Silvia Torres-Peimbert 
and Manuel Peimbert, at the occasion of a meeting in their honor. 
I am grateful to the Organizing Committee of 
for having made possible my participation to this stirry conference.


\end{document}